\documentstyle[12pt,epsf]{article}

\begin{document}

\pagestyle{empty}

\begin{flushright}
{\bf McGILL-01-20}\\
{\bf UA/NPPS-16-01}
\end{flushright}

\vglue 2cm

\begin{center} \begin{Large} \begin{bf}
Approximate Next-to-Leading Order and Next-to-Next-to-Leading Order
Corrections
\end{bf} \end{Large} \end{center}
\vglue 0.35cm
{\begin{center}
A.P.\ Contogouris$^{a,b,+}$ and
Z.\ Merebashvili$^{a,*,++}$,
\end{center}}
\parbox{6.4in}{\leftskip=1.0pc
{\it a.\ Department of Physics, McGill University, Montreal
H3A 2T8, Canada}\\
\vglue -0.25cm
{\it b.\ Nuclear and Particle Physics, University of Athens,
Athens 15771, Greece}
}
\begin{center}  
\vglue 1.0cm
\begin{bf} ABSTRACT \end{bf}
\end{center}
\vglue 1.0cm
{\rightskip=1.5pc 
\leftskip=1.5pc
\tenrm\baselineskip=12pt
 \noindent
For processes involving structure functions and/or fragmentation
functions, arguments that, over a range of a proper kinematic
variable, there is a part that dominates the next-to-leading order
(NLO) corrections are briefly reviewed. The arguments are tested
against more recent NLO and in particular complete
next-to-next-to-leading order (NNLO) calculations. A critical
examination of when these arguments may not be useful is also
presented.
}

\renewcommand{\thefootnote}{*}
\footnotetext{On leave from High Energy Physics Institute,
Tbilisi State University, University St. 9, 380086 Tbilisi,
Republic of Georgia.}
\renewcommand{\thefootnote}{+}
\footnotetext{Email: apcont@physics.mcgill.ca, acontog@cc.uoa.gr}
\renewcommand{\thefootnote}{++}
\footnotetext{Email: zaza@physics.mcgill.ca, mereb@sun20.hepi.edu.ge}

\newpage

\pagestyle{plain}
\setcounter{page}{1}

\renewcommand{\theequation}{1.\arabic{equation}}

\begin{center}\begin{large}\begin{bf}
1. INTRODUCTION
\end{bf}\end{large}\end{center}
\vglue .3cm

In Perturbative QCD there is now a great effort towards calculating
NNLO corrections $[$\ref{QCD1}-\ref{QCD3}$]$. One reason is that in
several cases the NLO corrections are found to be large. Other reasons
are that NNLO corrections are expected to increase the stability of
predicted cross sections against changes of schemes and scales and
that they will lead to more precise determinations of backgrounds
towards uncovering signals for new physics.

Although there is {\it no substitute} for a complete NNLO
calculation, since such calculations are in general very complicated,
as a {\it first step} one may try approximate ones. Such a step
has been presented in $[$\ref{apcsoft}$]$

Below we briefly review the arguments of $[$\ref{apcsoft}$]$. Sect.~2
mentions the results of certain more recent NLO calculations. Sect.~3
examines applications to the presently existing complete NNLO
calculations. Finally, Sect.~4, apart from certain other points,
discusses when approximate results may not be useful.

For processes involving structure functions and/or fragmentation
functions, in $[$\ref{apcsoft}$]$ it was argued that, over a range of
a proper kinematic variable, there is a part that dominates the
NLO; and this was used to explain the fact that, in a number of the
then existing NLO calculations, plotted against this kinematic
variable, in a wide range, the cross section was almost a constant
multiple of the Born.


To briefly review the essential ideas of $[$\ref{apcsoft}$]$, consider
the NLO contribution of the subprosses
$a(p_1)+b(p_2) \rightarrow \gamma (q)+d$ to the large-$p_T$ process
$A+B \rightarrow \gamma +X$:
\begin{eqnarray}
\nonumber
E \frac{d \sigma}{d^3 p} &=& \frac{\alpha_s (\mu )}{\pi } \sum_{a,b}
\int \frac{dx_a}{x_a} \frac{dx_b}{x_b} F_{a/A}(x_a, M)
F_{b/B}(x_b, M) \bigg[ \hat{\sigma} _B \delta \left(1+\frac{t+u}{s}
\right)  +  \\
\label{physical}
&&\frac{\alpha_s (\mu )}{\pi } f \theta
\left(1+\frac{t+u}{s}\right) \bigg] +
(1-\delta_{ab})(A\leftrightarrow B, \eta\leftrightarrow -\eta),
\end{eqnarray}
where $F_{a/A}$, $F_{b/B} $ are parton momentum distributions to the
hadrons $A,B$, $\mu$
and $M$ are the renormalization and factorization scales, $\eta$ the
c.m. pseudorapidity,
\[
s=(p_1+p_2)^2 , {\rm \hspace{.3in}} t=(q-p_1)^2 ,{\rm \hspace{.3in}} 
u=(q-p_2)^2
\]
and $\sigma_B$ and $f$ are functions of $s$, $t$, $u$ corresponding to
the Born and the higher order correction (HOC). Introducing the
dimensionless variables
\begin{equation}
v=1 +t/s ,{\rm \hspace{.6in}} w=-u/ (s+t)
\end{equation}
($s+t+u=sv(1-w)$), the HOC have the following overall structure:
\[
f ( v,w) =f_s ( v,w) +f_h ( v,w),
\]
where
\begin{eqnarray}
\nonumber
f_s(v,w) &=& a_1 (v) \delta (1-w) +b_1 (v) \frac{1}{(1-w)_+} +
c_1(v) \left( \frac{\ln (1-w) }{1-w} \right) _+   +   \\
&&\bigg( a_2(v) \delta (1-w) + b_2 (v) \frac{1}{(1-w)_+} \bigg) \ln
\frac{s}{M^2} + c_2(v) \delta(1-w) \ln\frac{s}{\mu^2},
\end{eqnarray}
where $1/(1-w)_+$ and $(\ln(1-w)/(1-w))_+$ are well known
distributions. The function $f_h(v,w)$ contains no distributions and,
in general, has a complicated analytic form.

Now denote by $\sigma_s$ and $\sigma_h$ the contributions of $f_s$ and
$f_h$ to $E d \sigma / d^3 p$ and consider the ratio
\begin{equation}
L=\sigma_h / (\sigma_s +\sigma_h) ;
\end{equation}
then, at sufficiently large $x_T$, for fixed total c.m. energy
$\sqrt{S}$, as $p_T$ (or $x_T \equiv 2p_T/ \sqrt{S}$) increases, $|L|$
decreases.

To see the reason, consider a plot of $x_b$ vs $x_a$ for $\eta=0$
(Fig.~1). The integration region in (\ref{physical}) is bounded
by $w=1$, $x_a=1$ and $x_b=1$ (hatched region). Now, for $x$ not too
small, $F_{a/A} (x, M)$ behaves like $(1-x)^n$; with
$A=$proton, $n$ is fairly large ($\geq 3$); also due to scale
violations, $n$ increases as $p_T$ increases.
Then contributions arising from the region away from $w=1$
are supressed by powers of $1-x_a$ and/or $1-x_b$.
Now, in $f_s$, the terms proportional to $\delta (1-w)$ contribute at
$w=1$ (and so does $\hat{\sigma}_B$) whereas the
rest give a contribution increasing as $w\rightarrow 1$. On the other
hand, the multitude of terms of $f_h$ contribute more or less
uniformly in the integration region $\theta(1-w)$ and their
contribution $\sigma _h$ is suppressed. As $x_T$ increases at fixed
$S$, the integration region shrinks towards $x_a =x_b =1$ and the
suppression of $\sigma _h$ increases.

The mechanism is tested by writing the distributions in the form
$[$\ref{apcsoft}(a)$]$
\begin{equation}
F_{a/A} (x, M)=F_{b/B} (x, M)=(1-x)^N
\end{equation}
and choosing a fictitious $N>>n$ or choosing $0<N<<n$. Then the ratio
$L$ with the first choice decreases faster and with the second choice
decreases slower then for $N=n$.

Next we neglect $f_h (v,w)$ and make the rough approximations
$1/(1-w)_+ \sim \delta (1-w)$,
$(\ln(1-w)/(1-w))_+ \sim \delta (1-w)$.
Furthermore, we note that $b_1(v), c_1(v), a_2(v), b_2(v), c_2(v)$ and
part of $a_1(v)$ are either proportional to the Born term or contain
the Born term times a smooth function of $v$; the rest of $a_1(v)$ is
also a smooth function of $v$ (see e.g. Eq.~(C.8) of
$[$\ref{apcsoft}(a)$]$ or Eq.~(4.11) of $[$\ref{apcsoft}(b)$]$). The
Born term itself is a smooth function of $v$. Thus as a first
approximation we write
\begin{equation}
f(v,w)\approx A\hat{\sigma}_B(v)\delta(1-w)
\end{equation}
where $A\approx const.$ This results in $Ed\sigma/d^3p$ of roughly the
same shape as $Ed\sigma_{\rm Born}/d^3p$

The same argument can be made in terms of the moments of the functions
$\delta(1-w), 1/(1-w)_+, (\ln(1-w)/(1-w))_+$ and of the functions in
$f_h(v,w)$ $[$\ref{apcsoft}(a)$]$. Clearly, $\sigma_s$ contains all the
soft, collinear and virtual contributions to $Ed\sigma/d^3p$.

At NLO the Bremsstrahlung (Brems) contributions to $f_s$ are
determined via simple formulae $[$\ref{apcsoft}$]$:
E.g. for $gq\rightarrow \gamma q$ the Brems contributions arise from
products of two graphs $gq\rightarrow \gamma qg$. If in both graphs
the emitted $g$ arises from initial partons (g or q), the contribution
in $d=4-2\varepsilon$ dimensions is
\begin{equation}
\label{init}
\frac{d\sigma _{\rm init}}{dvdw} \sim T_0 ^{(gq)} (v,\varepsilon ) N_c
\left( -\frac{2}{\varepsilon }\right)
\left( \frac{v}{1-v} \right) ^{-\varepsilon }
(1-w)^{-1-2\varepsilon } \left( 1+\varepsilon ^2
\frac{\pi ^2}{6} \right),
\end{equation}
where $T_0 ^{(gq)} (v,\varepsilon )$ is essentially the Born cross
section in $d$ dimensions. If in at least one of the graphs the emitted
$g$ arises from the final parton ($q$), then
\begin{equation}
\label{fin}
\frac{d\sigma_{\rm fin}}{dvdw} \sim T_0 ^{(gq)} (v,\varepsilon ) C_F
v^{-\varepsilon }
(1-w)^{-1-\varepsilon } \tilde P _{qq} (\varepsilon )
\end{equation}
where
\begin{equation}
\tilde P_{qq}(\varepsilon )=\frac{\Gamma(1-2\varepsilon )}{\Gamma ^2
(1-\varepsilon)}
\int _0 ^1 y^{-\varepsilon} (1-y)^{-\varepsilon } P_{qq}
(y,\varepsilon )
\end{equation}
and $P_{qq}(y,\varepsilon) =2/(1-y)-1-y-\varepsilon (1-y)$, the split
function in $n$ dimensions ($y<1$). Expanding
\begin{equation}
(1-w)^{-1-\varepsilon}=-\frac{1}{\varepsilon} \delta (1-w) +
\frac{1}{(1-w)_+} - \varepsilon \left( \frac{\ln(1-w)}{1-w} \right)_+
+{\cal O}(\varepsilon ^2)
\end{equation}
as well as $(v/(1-v))^{-\varepsilon }$ and  $v^{-\varepsilon }$ in
powers of $\varepsilon $ we determine the contributions. The singular
terms $\sim 1/\varepsilon ^2$ and $1/\varepsilon$ cancel by adding the
loop contributions and proper counterterms.

\renewcommand{\theequation}{2.\arabic{equation}}
\setcounter{equation}{0}
\vglue 1cm
\begin{center}\begin{large}\begin{bf}
2. FURTHER NLO CALCULATIONS
\end{bf}\end{large}\end{center}
\vglue .3cm

In addition to the examples presented in Refs. $[$\ref{apcsoft}$]$,
the following are some NLO studies supporting the ideas of Sect.~1:
\begin{itemize}
\item[(a)]
Large $p_T$ $W$ and $Z$ production in $p \bar p$ collisions
$[$\ref{gonsalves}$]$. At $\sqrt{S}=0.63$ and 1.8 TeV,
for $p_T \geq 80$ GeV the cross sections $d\sigma /dp_T ^2$ are also
almost a constant multiple of the LO (Figs.~7 and 8 of
$[$\ref{gonsalves}$]$).
\item[(b)]
The production of two isolated direct photons in $p\bar{p}$ collisions
$[$\ref{cdf}$]$. At $\sqrt{S}=1.8$ TeV, when the $p_T$ of each photon
exceeds 10~GeV the shape of the NLO QCD cross section $d\sigma/dp_T$
differs little from that of the Born (Fig.~2 of $[$\ref{cdf}$]$).
\end{itemize}

Regarding NLO results for polarized reactions we mention the following:  
\begin{itemize}
\item[(a)]
Polarized deep inelastic Compton scattering $[$\ref{cpt}$]$, in
particular the contribution of the subprocess $\vec \gamma \vec q
\rightarrow \gamma q$ to large $p_T$ direct photon production in
polarized $\gamma - p$ collisions ($\vec \gamma \vec p \rightarrow
\gamma +X$).
At $\sqrt{S}=27$ and 170 $GeV$, for $x_T \geq 0.15$, it is $|L|<0.28$
and for sufficiently large $x_T$, $L$ decreases as $x_T\rightarrow 1$
(Fig.~4 of Ref.~$[$\ref{cpt}$]$). Also, denoting by $\sigma^{(k)}$ the
${\cal O}(\alpha _s ^k )$, $k=0,1$, contributions of $\vec \gamma \vec
q \rightarrow \gamma q$ to $E d \sigma / d^3 p$, for
$0.2\leq x_T \leq 0.8$ the factor $K_{\gamma q}=(\sigma ^{(0)} +\sigma
^{(1)})/\sigma ^{(0)}$ is found to differ little from a constant.
\item[(b)]
Large $p_T$ direct $\gamma $ production in longitudinally polarized
hadron collisions $[$\ref{ckmt},\ref{gordon}$]$. Here of interest
are the ${\cal O}(\alpha _s ^k )$, $k=1,2$, contributions of the
subprocess $\vec g \vec q \rightarrow \gamma q$. As $x_T$
increases, the ratio $-\sigma _h /\sigma _s $ steadily decreases
(Fig.~10 of $[$\ref{ckmt}$]$). The factor
$K_{gq}=(\sigma ^{(1)} +\sigma ^{(2)})/\sigma ^{(1)}$ is not constant,
but increases moderately (Fig.~2 of $[$\ref{ckmt}$]$).
\item[(c)]
Lepton pair production by transversely polarized hadrons
$[$\ref{ckm},\ref{vogelsang}$]$. At fixed $S$, with increasing
$\sqrt{\tau }= M_{l^- l^+}/\sqrt{S}$, the ratio $\sigma _h /\sigma _s
$ is found again to decrease (Fig.~3 of $[$\ref{ckm}$]$).
Again, the $K$-factor is not constant, but increases moderately 
(Fig.~1 of $[$\ref{ckm}$]$). 
\end{itemize}

The considerations of Sect.~1 explain also the following fact: Taking
as example large $p_T$ $\vec p \vec p \rightarrow \gamma +X$, at NLO,
apart from the HOC of the dominant subprocess $\vec g \vec q
\rightarrow \gamma q$, there are contributions from the extra
subprocesses $\vec g \vec g \rightarrow q \bar q \gamma$, $\vec q
\vec q \rightarrow qq \gamma$ and $\vec q \vec {\bar q}\,' \rightarrow
q{\bar q}\,'\gamma$, where $q,{\bar q}\,'$ are either of different quark
flavor or of the same flavor but interacting via exchange of a gluon.
In general, these are found to be relatively small (Figs.~3, 4 and 5 of
$[$\ref{ckmt}$]$). The reason
is that the extra subprocesses possess no terms involving
distributions (no loops and vanishing contributions of the type
(\ref{init}) and (\ref{fin})).

\renewcommand{\theequation}{3.\arabic{equation}}
\setcounter{equation}{0}
\vglue 1cm
\begin{center}\begin{large}\begin{bf}
3. NNLO CALCULATIONS
\end{bf}\end{large}\end{center}
\vglue .3cm

NNLO calculations have been carried for Drell-Yan (DY) production of
lepton pairs, $W^{\pm }$ and $Z$, and for the deep inelastic
(DIS) structure functions $F_j (x, Q^2 )$, $j=1,2$ and the longitudinal
part. Now the parts involving distributions
contain also terms of the type $(\ln^i (1-w)/(1-w))_+ $, with $i=2$
and 3 and $w$ a proper dimensionless variable.
The subsequent calculations are carried using the updated 
$\overline{\rm MS}$ CTEQ5M1 set of $[$\ref{CTEQ5}$]$, one of the
most recent sets of NLO parton distributions $[$\ref{NNLO}$]$.
We present results for $\mu=M=\sqrt{Q^2}$.

Beginning with DY, we are interested in the process $p p
\rightarrow \gamma ^* +X \rightarrow l^+ l^- +X$ and
to the cross section
\begin{equation}
d\sigma (\tau , S) /dQ^2 \equiv \sigma (\tau , S) 
\end{equation}
where $\tau=Q^2/S$ with $\sqrt{S}$ the total c.m. energy of the
initial hadrons and $\sqrt{Q^2 }$ the $\gamma^*$ mass
$[$\ref{matsuura},\ref{rijken}$]$. Here we deal with the subprocess
$q+\bar q\rightarrow \gamma ^*$ and its NLO and NNLO corrections
$[$\ref{matsuura}$]$. For DY, $w\sim \tau $. We use number of flavors
$n_f=4$.

Denote by $\sigma ^{(k)} (\tau , S)$, $k=0,1,2$, the ${\cal O}(\alpha
_s ^k )$ part of $\sigma (\tau , S)$, by $\sigma _s ^{(k)}$
the part of $\sigma ^{(k)}$ arising from distributions and by
$\sigma _h ^{(k)}$ the rest. Defining
\begin{equation}
L^{(k)}(\tau , S) = \sigma _h ^{(k)} (\tau , S) /  \sigma ^{(k)}
(\tau , S) 
\end{equation}
Fig.~2 shows $L^{(k)}$, $k=1,2$, as functions of $\tau $ for
$\sqrt{S}=20$ GeV. Clearly, for $\tau >0.3$: $L^{(1)} \leq 0.17$
and $L^{(2)} \leq 0.33$.

It is of interest also to see the percentage of $\sigma _h ^{(k)}$ of
the total cross section determined up to ${\cal O}(\alpha _s ^k )$.
Fig.~2 also shows the ratios $\sigma _h ^{(1)}/(\sigma ^{(0)} +\sigma
^{(1)} )$ and $\sigma _h ^{(2)}/(\sigma ^{(0)} +\sigma ^{(1)} +\sigma
^{(2)} )$ for the same $\sqrt{S}$; clearly, for $\tau \geq 0.2$
both ratios are less than 0.1.  

Now we turn to DIS $[$\ref{neerven},\ref{zijlstra}$]$. Here we deal
with the sum
\begin{equation}
\Sigma (x,Q^2)=u_v(x,Q^2)+d_v(x,Q^2),
\end{equation}
where $u_v$ and $d_v$ are the $u$-valence and $d$-valence quark
distributions in the proton.  We will deal with the subprocess 
$q+\gamma^* \rightarrow q$ and the nonsinglet part of its NLO and NNLO
corrections $[$\ref{neerven}$]$. For DIS, $w\sim x$.

Denote by $\Sigma^{(k)} (x,Q^2 )$, $k=0,1,2$, the ${\cal 
O}(\alpha_s^k)$ contribution, by $\Sigma_s ^{(k)}$ the part of
$\Sigma^{(k)}$ arising from distributions and by $\Sigma_h ^{(k)}$ the
rest. Defining
\begin{equation}
L^{(k)} (x,Q^2 ) = \Sigma_h^{(k)}(x,Q^2 )/ \Sigma^{(k)}(x,Q^2 )
\end{equation}
Fig.~3 presents $L^{(k)}(x,Q^2 )$, $k=1,2$, as functions of $x$ for
$\sqrt{Q^2}=5$ GeV. Now, for $x\leq 0.5$ $L^{(1)}$ is not
small, but this is due to the fact that $\Sigma_s ^{(1)}$ changes sign
and $\Sigma_h ^{(1)}$ stays $>0$, so at $x\approx 0.3$, $\Sigma^{(1)}$
vanishes. On the other hand, at $x\geq 0.6$, $L ^{(2)}$ is less than
0.28.

Fig.~3 also shows the ratios
$\Sigma_h^{(1)}/(\Sigma^{(0)}+\Sigma^{(1)} )$ and
$\Sigma_h^{(2)}/(\Sigma^{(0)}+\Sigma^{(1)}+\Sigma^{(2)} )$ for the
same $Q^2$; for $x \geq 0.3$ both ratios are less than 0.1.

The effect of neglecting $\sigma _h ^{(k)}$ in DY or $\Sigma_h^{(k)}$
in DIS is shown in Fig.~4. In DY, denoting  
\begin{eqnarray}
\nonumber
K_s&=&(\sigma^{(0)}+\sigma_s^{(1)}+\sigma_s^{(2)} )/\sigma^{(0)} \\
K&=&(\sigma^{(0)}+\sigma^{(1)}+\sigma^{(2)})/\sigma^{(0)}
\end{eqnarray}
we show $K_s (K)$ by solid (dashed) line at $\sqrt{S}=20$ GeV (upper
part). Clearly, as $\tau \rightarrow 1$,
$K_s \rightarrow K$, and for $\tau > 0.3$ the error is less than
$14\%$.
In DIS, denoting
\begin{eqnarray}
\nonumber
K_s&=&(\Sigma^{(0)}+\Sigma_s^{(1)}+\Sigma_s^{(2)} )/\Sigma^{(0)}  \\
K&=&(\Sigma^{(0)}+\Sigma^{(1)}+\Sigma^{(2)})/\Sigma^{(0)},
\end{eqnarray}
we show $K_s$ and $K$ at $\sqrt{Q^2}=5$ GeV (lower part). Again, as $x
\rightarrow 1$, $K_s \rightarrow K$. Now, in spite of the fact that
$L^{(k)}$ is, in general, not small, $K_s$ differs from $K$ even
less. The reason is that the NLO and NNLO corrections are smaller than
in DY, and so are $\Sigma_s ^{(k)}/\Sigma^{(0)}$.

\vglue 1cm
\begin{center}\begin{large}\begin{bf}
4. CONCLUDING REMARKS
\end{bf}\end{large}\end{center}
\vglue .3cm

The above discussion and examples show that for processes involving
structure and/or fragmentation functions, for not too small values of
a proper kinematic variable ($x_T$ for large-$p_T$ reactions, $\tau $
for DY, $x$ for DIS), one may retain only that part of the differential
cross section arising from distributions $[$\ref{r18}$]$. At NNLO the
range of this variable is larger for DY than for DIS (Figs.~2 and 3).
Yet, regarding the $K$-factor, which determines the physically important
quantity, DIS is somewhat advantageous (Fig.~4).

As we go to NNLO, in view of the presence of terms of the type
$(\ln^i (1-w)/(1-w))_+$ with $i=2$ and 3, the rough approximation of
replacing $(\ln(1-w)/(1-w))_+$ by $\delta(1-w)$ is becoming worse. In
general, this implies that with NNLO corrections, the shape of a physical
quantity should deviate more than that of the Born term.

It is desirable (and nontrivial) to extend the formulas (\ref{init}) and
(\ref{fin}) to the NNLO and perhaps even higher orders.

The question now is when the arguments of Sec.~1 may not be useful. Such a
case is when, over a wide range of $w$, $\sigma_h^{(k)}$ is comparable and
of opposite sign to $\sigma_s^{(k)}$. Then $\sigma_s^{(k)} +
\sigma_h^{(k)}$ is small and $L^{(k)}$ is large in absolute value. Even
then, for $w$ very near $1$, $|L^{(k)}|$ should decrease, but in that case
threshold resummation $[$\ref{resum}$,$\ref{nnloln}$]$ is important, and
the approximation is not useful. Of course, in such a case, the correction
$|\sigma^{(k)}| = |\sigma_s^{(k)} + \sigma_h^{(k)}|$ will be small. The
point, however, is that we do not see how one can determine such a case
without calculating $\sigma_h^{(k)}$.

\vglue 1cm
\begin{center}\begin{large}\begin{bf}
ACKNOWLEDGEMENTS
\end{bf}\end{large}\end{center}
\vglue .3cm 

We would like to thank E.~Basea and G.~Grispos for checking certain of
our results. The work was also supported by the Natural Sciences and
Engineering Research Council of Canada and by the Secretariat for
Research and Technology of Greece.

\vglue 1cm
\begin{center}\begin{large}\begin{bf}
REFERENCES
\end{bf}\end{large}\end{center}
\vglue .3cm
                        
   \begin{list}{$[$\arabic{enumi}$]$}
    {\usecounter{enumi} \setlength{\parsep}{0pt}
     \setlength{\itemsep}{3pt} \settowidth{\labelwidth}{(99)}
     \sloppy}
\item \label{QCD1}
QCD, hep-ph/0005025.
\item \label{QCD2}
The QCD and the Standard Model Working Group, hep-ph/0005114.
\item \label{QCD3}
Parton Distributions Working Group, hep-ph/0006300.
\item \label{apcsoft}
(a) A.~P.~Contogouris, N.~Mebarki and S.~Papadopoulos,
Intern. J. Mod. Phys. {\bf A5}, 1951 (1990);
(b)~A.~P.~Contogouris and S.~Papadopoulos,
Mod. Phys. Lett. {\bf A5}, 901 (1990).
\item \label{gonsalves}
R.~Gonsalves, J.~Pawlowski and C.~F.~Wai,
Phys. Rev. {\bf D40}, 2245 (1989).
\item \label{cdf}
CDF Collaboration, F.~Abe et al, Phys. Rev. Lett. {\bf 70}, 2232
(1993).
\item \label{cpt}
A.~P.~Contogouris, S.~Papadopoulos and F.~V.~Tkachov,
Phys. Rev. {\bf D46}, 2846 (1992).
\item \label{ckmt}
A.~P.~Contogouris, B.~Kamal, Z.~Merebashvili and F.~V.~Tkachov,
Phys. Rev. {\bf D48}, 4092 (1993); {\bf D54}, 701 (1996) (Erratum).
\item \label{gordon}
L.~Gordon and W.~Vogelsang, Phys. Rev. {\bf D49}, 70 (1994).
\item \label{ckm}
A.~P.~Contogouris, B.~Kamal and Z.~Merebashvili,
Phys. Lett. {\bf B337}, 169 (1994).
\item \label{vogelsang}
W.~Vogelsang and A.~Weber, Phys. Rev. {\bf D48}, 2073 (1993).
\item \label{CTEQ5}
H.~Lai et al., Eur. Phys. J. {\bf C12}, 375 (2000).
\item \label{NNLO}
Exact NNLO sets are not yet available. See e.g. M.~Grazzini,
hep-ph/0105299.
\item \label{matsuura}
T.~Matsuura, S.~van~der~Marck and W.~van~Neerven,
Phys. Lett. {\bf B211}, 171 (1988); Nucl. Phys. {\bf B319}, 570
(1989); R.~Hamberg, W.~van~Neerven and T.~Matsuura,
Nucl. Phys. {\bf B359}, 343 (1991).
\item \label{rijken}
P.~Rijken and W.~van~Neerven, Phys. Rev. {\bf D51}, 44 (1995).
\item \label{neerven}
W.~van~Neerven and E.~Zijlstra, Phys. Lett. {\bf B272}, 127 (1991).
\item \label{zijlstra}
E.~Zijlstra and W.~van~Neerven,
Phys. Lett. {\bf B273}, 476 (1991); Nucl. Phys. {\bf B383}, 525
(1992).
\item \label{r18}
At NLO, certain of the terms of $f_h(v,w)$ are also given by simple
analytic formulas similar to (\ref{init}) and (\ref{fin}). See
S.~Papadopoulos, Ph.D. Thesis (McGill Univ. 1989).
\item \label{resum}
E.~Laenen, G.~Sterman and W.~Vogelsang, Phys. Rev. {\bf D63}, 114018 
(2001); G.~Sterman and W.~Vogelsang, Journ. H.E.P. {\bf 02}, 016
(2001); and references therein.
\item \label{nnloln}
A.~Vogt, Phys. Lett. {\bf B497}, 228 (2001).
\end{list}


\vglue 1cm
\begin{center}\begin{large}\begin{bf}
FIGURE CAPTIONS
\end{bf}\end{large}\end{center}
\vglue .3cm

   \begin{list}{Fig.~\arabic{enumi}} 
    {\usecounter{enumi} \setlength{\parsep}{0pt}
     \setlength{\itemsep}{3pt} \settowidth{\labelwidth}{(99999)}
     \sloppy}
\item
The integration region in the expression (\ref{physical}) for c.m.
pseudorapidity $\eta=0$.
\item
The ratios $L^{(2)}=\sigma_h^{(2)}/\sigma^{(2)}$ and $\sigma_h^{(2)}/
(\sigma^{(0)}+\sigma^{(1)}+\sigma^{(2)})$ (solid lines) as well as
$L^{(1)}=\sigma_h^{(1)}/\sigma^{(1)}$ and $\sigma_h^{(1)}/
(\sigma^{(0)}+\sigma^{(1)})$ (dash-dotted lines) for Drell-Yan
lepton-pair production versus $\tau=Q^2/S$ at $\sqrt{S}=20$ GeV.
\item
The ratios $L^{(2)}=\Sigma_h^{(2)}/\Sigma^{(2)}$ and $\Sigma_h^{(2)}/
(\Sigma^{(0)}+\Sigma^{(1)}+\Sigma^{(2)})$ (solid lines), where $\Sigma
\equiv u_v+d_v$, as well as
$L^{(1)}=\Sigma_h^{(1)}/\Sigma^{(1)}$ and $\Sigma_h^{(1)}/
(\Sigma^{(0)}+\Sigma^{(1)})$ (dash-dotted lines) for $q+\gamma^*
\rightarrow q$ versus $x$ at $Q^2=25$ GeV$^2$.
\item
$K$-factors: Approximate $K_s$ (solid lines) and exact $K$ (dotted
lines) for the Drell-Yan case of Fig.~2 (upper part) and for the case of
Fig.~3 (lower part).

\end{list}

\newpage
\pagestyle{empty}

\centering
\mbox{\epsfysize=200mm\epsffile{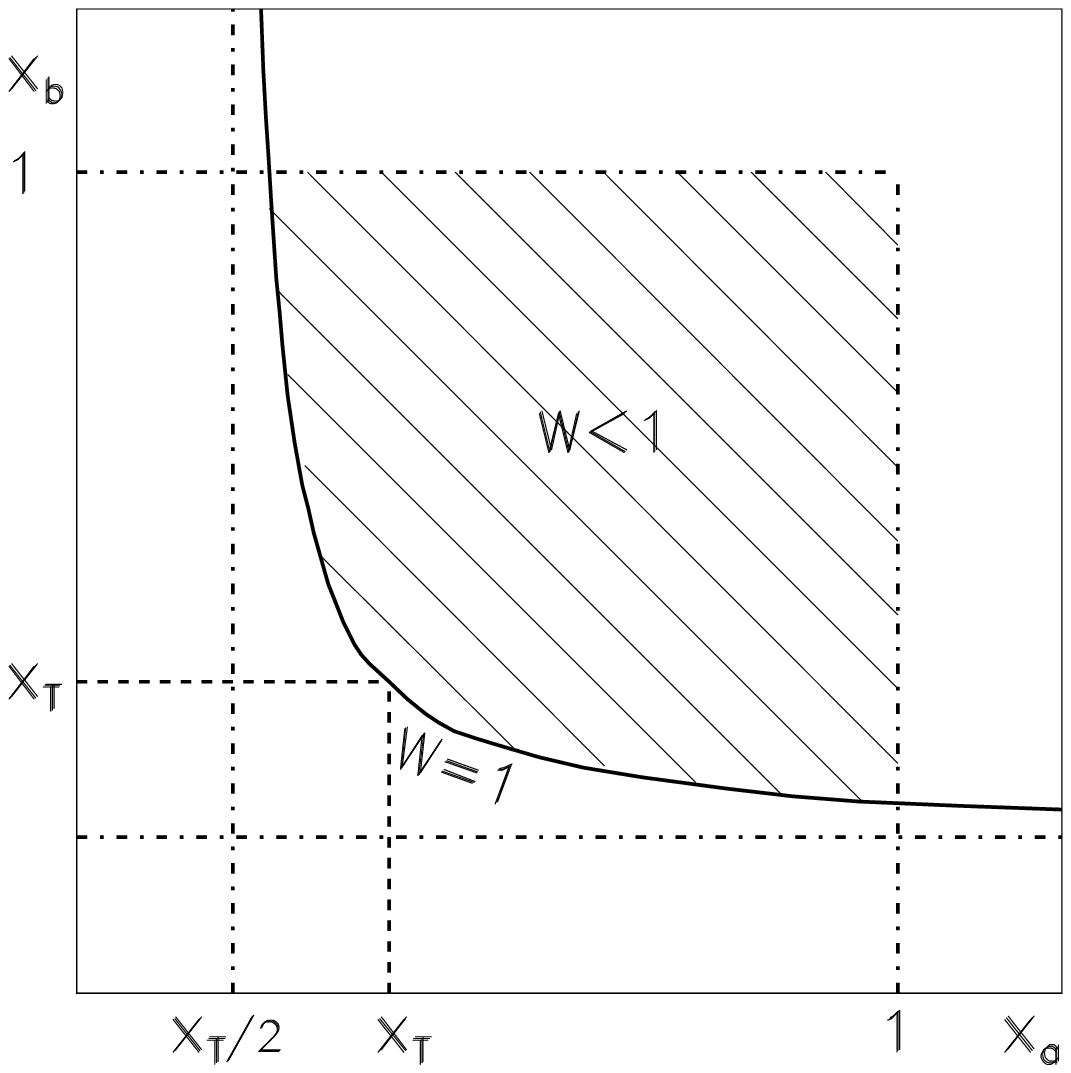}}

\begin{center}
\vglue -1 cm
\large{Fig.~1}
\end{center}

\newpage
\centering
\mbox{\epsfysize=200mm\epsffile{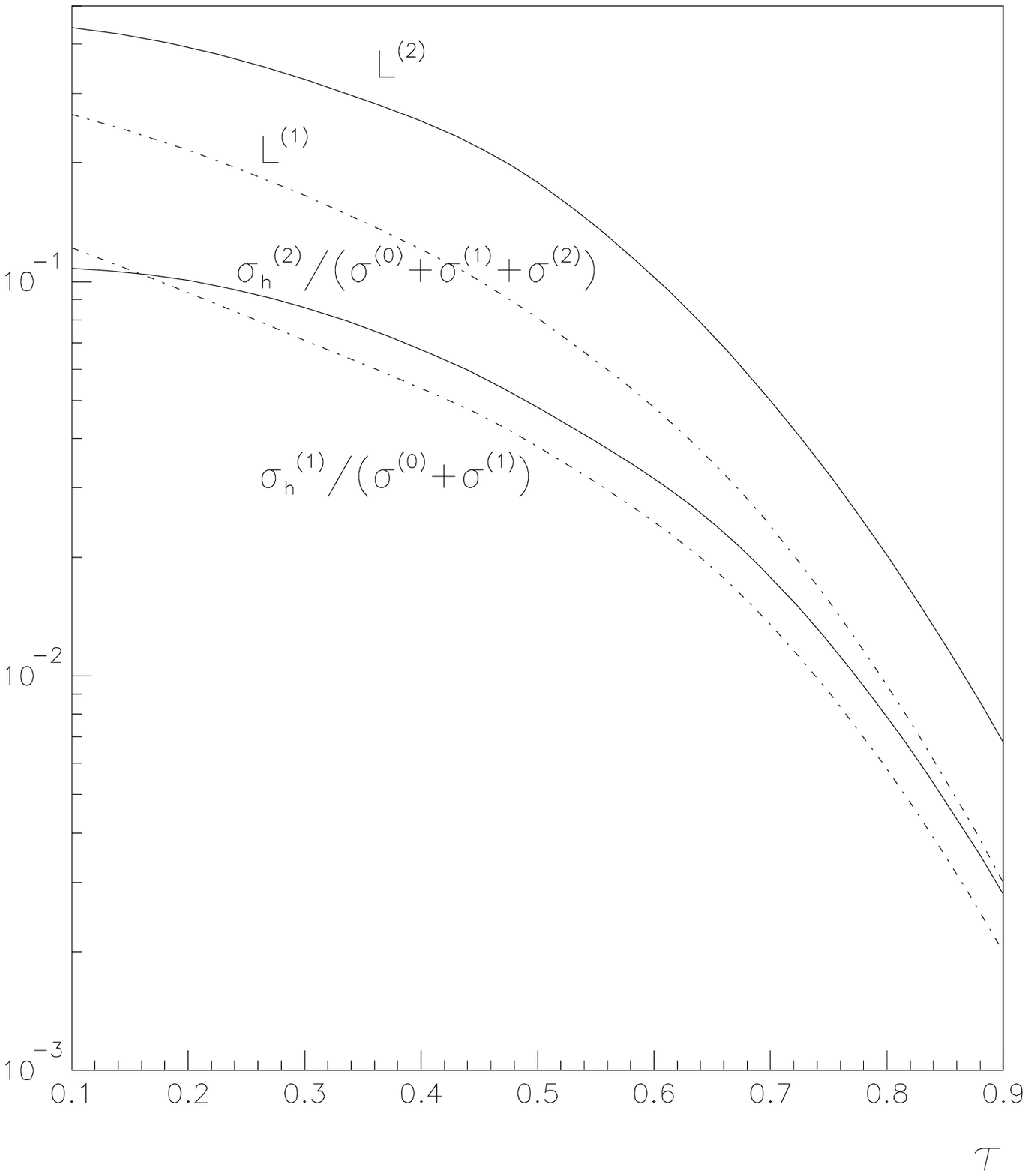}}

\begin{center}
\vglue -1 cm  
\large{Fig.~2}
\end{center}

\newpage

\centering  
\mbox{\epsfysize=200mm\epsffile{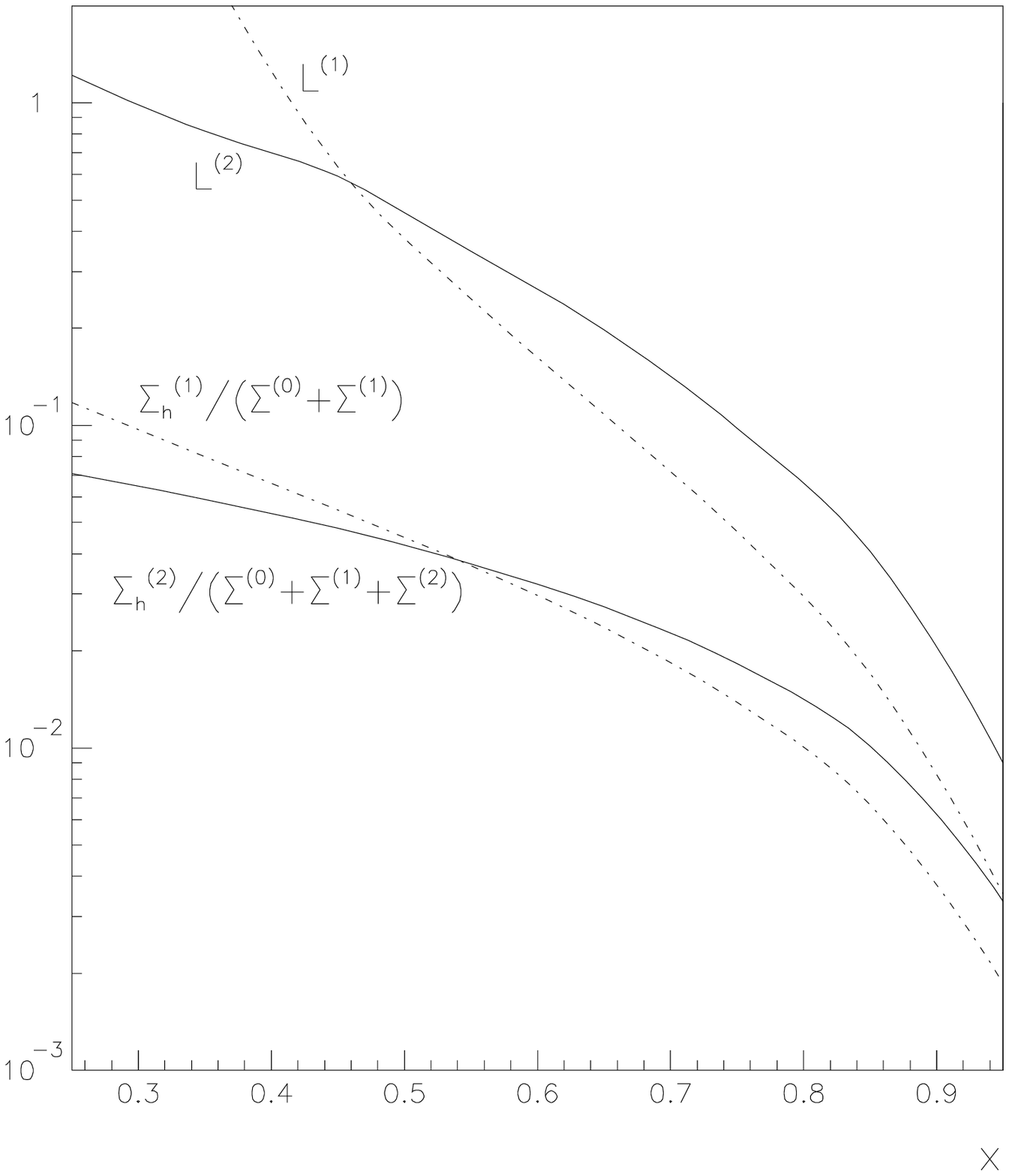}}

\begin{center}
\vglue -1 cm
\large{Fig.~3}
\end{center}

\newpage
\centering
\mbox{\epsfysize=200mm\epsffile{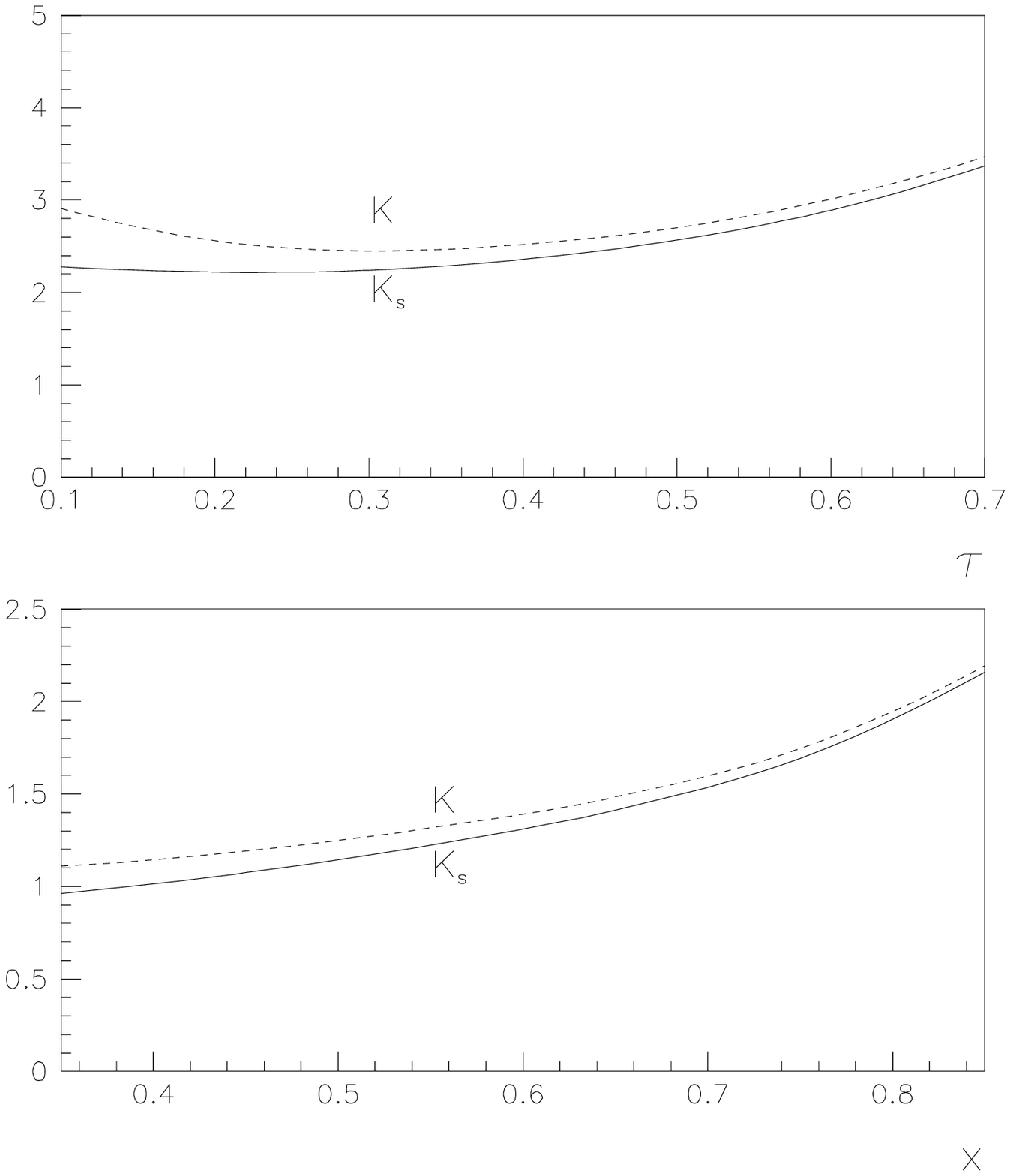}}

\begin{center}
\vglue -1 cm  
\large{Fig.~4}
\end{center}

\end{document}